\documentclass[12pt]{iopart}

\usepackage{iopams}
\usepackage{cite}
\usepackage{graphicx}
\usepackage{booktabs}
\usepackage[colorlinks,
            linkcolor=red,
            anchorcolor=blue,
            citecolor=green
            ]{hyperref}

\begin{document}

\title[Inter-Spacecraft Tilt-to-Length Noise Reduction Algorithm for  Taiji Mission]{Inter-Spacecraft Tilt-to-Length Noise Reduction Algorithm for Taiji Mission}

\author{Qiong Deng$^1$, Leiqiao Ye$^2$, Ke An$^{3,4}$, Yidi Fan$^{3,4}$, Ruihong Gao$^5$, Ziren Luo$^{2,5,6}$, Minghui Du$^{5,*}$, Pengcheng Wang$^{3,4,*}$, and Peng Xu$^{1,2,5,6,*}$}

\address{
$^1$ Lanzhou Center of Theoretical Physics, Lanzhou University, Lanzhou 730000, China\\
$^2$ Key Laboratory of Gravitational Wave Precision Measurement of Zhejiang Province, Hangzhou Institute for Advanced Study, UCAS, Hangzhou,  310024, China\\
$^3$ Innovation Academy for Microsatellites of Chinese Academy of Sciences, Shanghai, 201304, China\\
$^4$ Key Laboratory for Satellite Digitalization Technology, Chinese Academy of Sciences, Shanghai, 201210, China\\
$^5$ Center for Gravitational Wave Experiment, National Microgravity Laboratory, Institute of Mechanics, Chinese Academy of Sciences, Beijing 100190, China\\ 
$^6$ Taiji Laboratory for Gravitational Wave Universe (Beijing/Hangzhou), University of Chinese Academy of Sciences (UCAS), Beijing 100049, China\\ 
}

\eads{\mailto{duminghui@imech.ac.cn}, \mailto{wangpc@microsate.com}, \mailto{xupeng@imech.ac.cn}}


\begin{abstract}
The Taiji mission for space-based gravitational wave (GW) detection employs  laser interferometry to measure picometer-scale distance variations induced by GWs. 
The tilt-to-length (TTL) coupling noise in the inter-spacecraft interferometers, 
which originates from the angular jitters of the spacecrafts and the movable optical subassemblies, 
is predicted to be one of the main noise sources that might  reduce Taiji's sensitivity to GWs.
Since these angular jitters can be read out through the differential wavefront sensors, it is possible to suppress TTL noise during the data processing stage by fitting and subtracting it.
This paper proposes an improved algorithm for TTL noise suppression,  which addresses the issue of unknown noise floor required for optimal estimation in the practical detection scenario, and the design of this algorithm takes into account the presence of GW signals.
The algorithm is validated via numerical simulation, which is built on a spacecraft dynamics simulation incorporating Taiji's drag-free and attitude control system. 
We also demonstrate the robustness of this algorithm by varying TTL coefficients at different levels, indicating that our algorithm is applicable to a range of payload statuses, and ultimately providing a critical advancement toward realizing Taiji's full sensitivity.
\end{abstract}

%
%
%
%
%

\section{Introduction\label{sec:I}}

After ten years after the first direct detection of gravitational wave (GW) signal by LIGO-Virgo in 2015~\cite{PhysRevLett.116.061102},  more than a hundred events have been observed by the LIGO-Virgo-KAGRA collaboration,  and GW detections, as a complementary method to traditional electromagnetic observations, become a new observational window to the Universe.
To observe heavier GW sources or sources of low frequencies ($\sim$ miliHertz band), the Laser Interferometer Space Antenna (LISA) mission, led by the European Space Agency (ESA), was proposed in the 1990s \cite{Karsten_Danzmann_1996,2017arXiv170200786A}.
With similar design concepts and architecture, different LISA-like space-based GW antennas were suggested and are under active studies \cite{hu2017taiji,luo2021taiji,Luo_2016,AstrodGW,LISAmax}.  
Among these, the China-led projects, including the Taiji \cite{hu2017taiji,luo2021taiji} and TianQin \cite{Luo_2016} missions, is currently in a substantial research and development phase.

The Taiji mission, proposed by Chinese Academy of Sciences since 2016 \cite{cyranoski2016chinese}, is a heliocentric mission aiming at miliHertz band sources including massive black hole binary mergers, extreme mass-ratio inspirals, galactic binaries, stochastic background, etc. 
The science operations of LISA, Taiji and TianQin may overlap in the 2030s, and such antenna networks could significantly improve the detection capabilities for certain sources \cite{CAI20241072,Ruan:2020smc,PhysRevD.104.024012}.
With design variations compared to the current LISA mission concept, the Taiji mission will launch three space-crafts (SC) to form an equilateral triangle constellation with armlength $\sim3\times 10^6$ km and $\sim 20^\circ$ following or ahead of Earth \cite{hu2017taiji,ziren2020introduction,gong2015descope,guo2010space}. 
Each SC is equipped with two movable optical subassemblies (MOSAs), and each MOSA consists mainly of a telescope, an optical bench (OB), and a gravitational reference sensor (GRS).
A test mass (TM), as the free-falling reference and the end mirror of inter-satellite interferometers, will be suspended and floating within each GRS by means of the electrostatic compensation forces on the TM itself and the drag-free controls of the SC. 
Incident GWs induce tidal accelerations among these free-falling TMs and leave measurable signals in the interferometric readouts \cite{cai2017gravitational,huang2017gravitational,ruan2020taiji,zhao2020prospects,luo2020brief,luo2021taiji,luo2022recent,zhao2022preliminary,zhang2019overall}.

For such a measurement scheme of LISA and Taiji missions, the most important noise is from the frequency instabilities of the onboard lasers, which could be $7\sim 8$ orders of magnitude larger than the required resolution level \cite{tinto1999cancellation,armstrong1999time}.   
With the time delay interferometer (TDI) techniques, such key noise can be efficiently suppressed and resolved in the data post-processing stage \cite{tinto1999cancellation,armstrong1999time,staab2024laser,bayle2021adapting,vallisneri2021time,nam2022tdi,tinto2021time}. 
After the suppression of the laser frequency noise, a major one among the next leading noise sources is the tilt-to-length (TTL) coupling, which had been measured and verified in the LISA PathFinder (LPF) mission \cite{Hartig:2023ofu,PhysRevD.108.102003}.
During the science operations of LISA and Taiji, the drag-free and attitude control system (DFACS) will take care of up to 60 degrees of freedom of the 3 SCs, 6 MOSAs, and 6 TMs to maintain the inter-satellite laser links of the constellations and the free-falls of the TMs along the sensitive directions, 
which also leave with the residual attitude jitters of SCs, MOSAs, and TMs. 
Such jitters will produce the TTL coupling displacement noises in the interferometric length measurements, as  reviewed by \emph{e.g.} Ref.~\cite{wanner2024depth}.

Based on recent studies, one expects that it is challenging to suppress sufficiently the TTL noise to the required noise level ($<{\rm pm/Hz^{1/2}}$) by hardware designs and by fine-tuning of the alignments of interferometers before or after launch  ~\cite{armano2019lisa,armano2016sub,McNamara_2008,armano2018calibrating,armano2020lisa,hartig2022geometric,tao2024approximate}.
Therefore,  the fits and subtractions of the residual TTL noises in the post-processing stage will be necessary.  
Considered that the magnitudes of the residual jitters are rather small ($\sim 10\ {\rm nrad/Hz^{1/2}}$), the TTL noise can be well approximated by a linear model proportional to the relative attitude jitter angle, with higher-order terms neglected.

The current strategy for suppressing TTL noise through data processing follows a three-step approach~\cite{paczkowski2022postprocessing,george2023calculating,hartig2024post,armano2023tilt,houba2023time,zhao2020tilt}. 
First, the contributions of TTL noise to TDI observables are modeled using a first-order (linear) approximation, namely the extra optical path variations are proportional to the angular jitters of MOSAs. 
Second,  a likelihood function (or a loss function) is established by comparing the model to TDI data. 
Finally, the TTL coupling coefficients are estimated by maximizing this likelihood function (or by minimizing the loss function). 
The fitted TTL noise is then subtracted from the TDI data, thereby achieving effective noise suppression. 
However, the aforementioned method has certain limitations. 
Firstly, according to the principle of matched filtering~\cite{matched_filter_Davis1989ARO,matched_filter_Jaranowski_Krolak_2009}, 
the power spectral density (PSD) of the noise floor (i.e. the so-called secondary noises, which are the residual noises  after TTL noise has been fully subtracted, mainly including the optical metrology system (OMS) noises of interferometers and acceleration (ACC) noises of TMs) is a necessary element of the likelihood function. 
Given that  the amplitude of TTL noise is half or one order of magnitude larger than that of the noise floor, it might  not be feasible to assume that we already have the knowledge about this PSD 
before the subtraction of  TTL noise. 
Secondly, the TDI data contains not only TTL and other noises, but also GW signals. 
Therefore, the impact of GW signals cannot be ignored. 
In fact, Ref.~\cite{littenberg2023prototype} claims that the estimation of TTL coefficients  needs to be incorporated into the ``global fitting'' of numerous GW source and noise parameters. 
A recent study in Ref.~\cite{Hartig_2025} showed that the bright merger signal from $10^6  M_{\bigodot}$ massive black hole binary (MBHB)  may  induce a 0.33 mm/rad bias in TTL coefficient estimation. 
To complement,  we further investigate the impact of a $5\times 10^4  M_{\bigodot}$ MBHB, whose merger frequency is relatively higher and nearer to the frequency range where  TTL noise is  most pronounced in our simulation.
At last, the drag-free attitude control system (DFACS) controls the SCs to follow the TMs to perform inertial motion  along the sensitive  (\emph{i.e.} the axes of inter-spacecraft laser links).
The angular jitters of the SCs and MOSAs are essentially the residual noise of drag-free control,  
therefore the study of practical TTL noise subtraction algorithms should, in principle, be based on the simulations of DFACS~\cite{robert2022microscope,armano2018beyond}. 

This paper proposes and validates an enhanced method for suppressing TTL noise, seeking  to address the aforementioned limitations. 
In Section~\ref{sec:II}, we present the TTL noise model for the Taiji project, along with a detailed description on data simulation. 
The design and theoretical analysis of our algorithm are presented in Section~\ref{sec:III}, which is followed by the results based on simulation in Section~\ref{sec:IV}. 
After subtracting the TTL noise, the results are  shown to meet the noise requirements of Taiji. 
Finally, in Section~\ref{sec:V}, we draw the concluding remarks of this study.

\section{Model and simulation for TTL noise \label{sec:II}}
\subsection{TTL noise model \label{subsec:ttl_model}}

\begin{figure}
\centering
\includegraphics[width=0.5\textwidth]{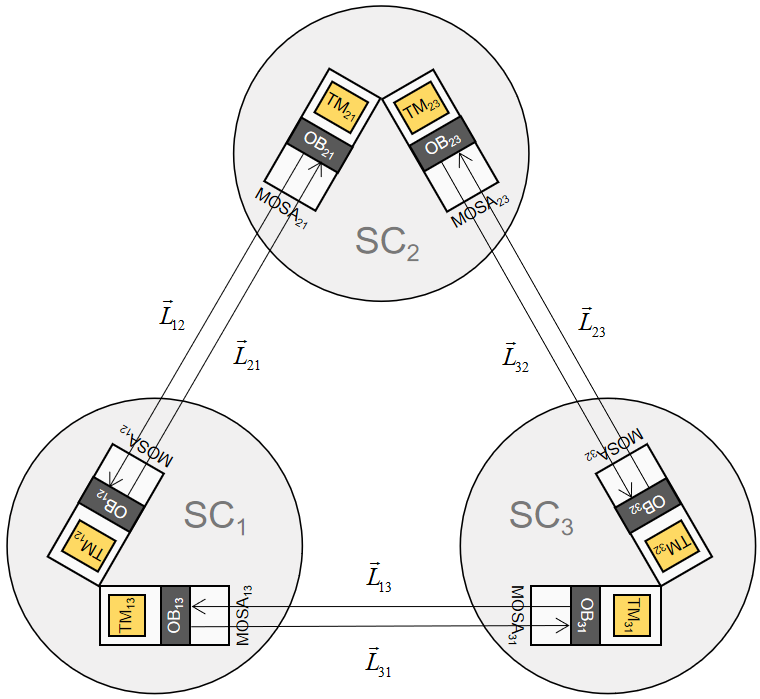}
\caption{Schematic of Taiji's spacecrafts (SCs), showing the indexing conventions of  movable optical subassemblies (MOSAs), optical benches (OBs), test masses (TMs) and laser links. }
\label{卫星示意图} 
\end{figure}

The schematic of Taiji constellation is depicted in Figure~\ref{卫星示意图}, which includes three SCs, each equipped with two MOSAs. 
Each MOSA consists of a telescope, an OB, and a TM.
This paper uses the following indexing convention. The three SCs are arranged clockwise as SC$_1$, SC$_2$, and SC$_3$. Each MOSA is labelled with two indices $ij$, where $i$ represents the SC on which the MOSA is located and $j$ represents the SC with which the laser beam is exchanged. Subsystems contained within the MOSA and phasemeter readouts share the same indices as the MOSA does. It should be noted that for the arm $\vec L_{ij}$, the indexing convention differs from that of the MOSAs. 
Specifically, the first index $i$ denotes the SC that receives the laser, while the second index $j$ refers to the SC that emits the laser.
Our modeling of the SCs and MOSAs are in consistency with the setup of of DFACS simulation in Ref.~\cite{qianjiao2024design}, where  the left MOSAs are fixed to the SCs, and  the right ones can rotate around the $z$-axes of SCs.

\begin{figure}
\centering
\includegraphics[width=0.5\textwidth]{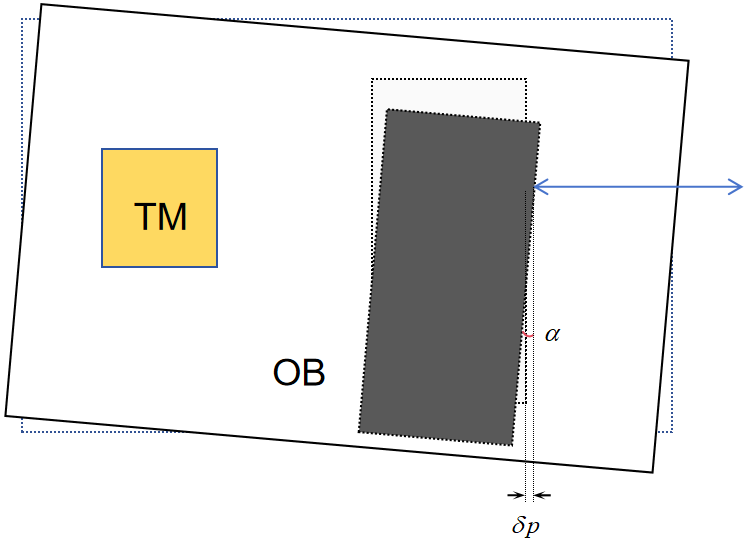}
\caption{An illustration for the geometric TTL coupling.}
\label{TTL示例图} 
\end{figure}

Along the nominal orbits, the attitudes of the SCs, MOSAs and TMs will inevitably experience attitude jitters. 
The DFACS will control the SCs to follow the inertial motions of the onboard TMs along their sensitive axes. 
However, limited by the precision of the drag-free control, the SCs and MOSAs will still have residual angular jitters relative to the corresponding lines of sight. These jitters will generate additional optical path noises in the laser interferometric measurements. 
For illustration, one may consider the example  in Figure~\ref{TTL示例图}. 
The angular jitter of the MOSA causes the optical bench to be non-perpendicular to the incident light, resulting in an additional optical path difference. 
Other mechanics of TTL coupling are also reported in the literature, such as the angular transmitter jitter coupling due to wavefront errors as described in ~\cite{wanner2024depth}. 
The TTL noises in test-mass interferometers (TMIs) are considered to be small enough such that it can be safely ignored~\cite{wanner2024depth,shen2024suppression}. 
In this paper, the TTL noise under considerations refers to the inter-spacecraft optical path length variations caused by attitude jitters of the MOSAs and the SCs.

\begin{figure}
\centering
\includegraphics[width=0.5\textwidth]{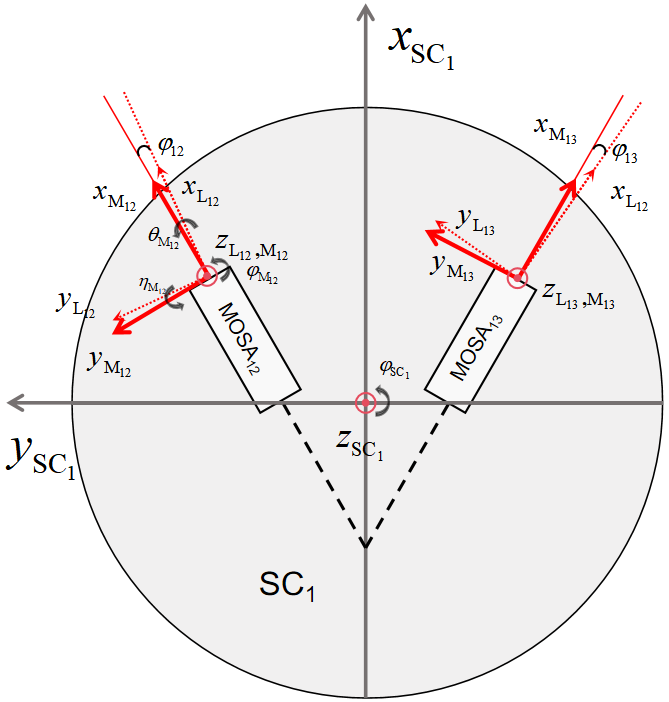}
\caption{The coordinate system of SC$_1$ and the two MOSAs mounted on it.}
\label{新坐标} 
\end{figure}

To better describe how TTL noise affects the measurements of inter-spacecraft interferometers (ISIs),  we establish coordinate systems for each SC and MOSA. 
For example, Figure~\ref{新坐标} illustrates the coordinate systems of SC$_1$ and the two MOSAs mounted on it. 
For each SC, 
the $xy$-plane is spanned by the axes of two telescopes, 
and the $z$-axis is normal to this plane, pointing outward. 
Each MOSA$_{ij}$ is associated to 2 coordinate systems dubbed ${\rm M}_{ij}$ (${\rm M}$ for MOSA) and ${\rm L}_{ij}$ (${\rm L}$ for laser), respectively.
The ${\rm M}_{ij}$ frame is fixed to  MOSA$_{ij}$ and hence jitters together with the MOSA, with the $x$-axis pointing along the axis of telescope, the $z$-axis parallel to the SC's, and the  $y$-axes defined so that $\{x, y, z\}$ axes form a right-handed coordinate system. 
Conversely, the  $x$-axis of  ${\rm L}_{ij}$ frame is perfectly aligned with the incident laser, therefore $\hat{x}_{{\rm L}_{ij}}$ represents the ``nominal'' orientation of the MOSA. 
The two lasers received by SC$_i$ span a common $xy$-plane for the two corresponding ${\rm L}_{ij}$ frames, hence determining a common $z$-axis. 
The roll $\theta_{ij}$, pitch $\eta_{ij}$ and yaw $\phi_{ij}$ represent the rotations of $\hat{x}_{{\rm L}_{ij}}$ (laser) relative to ${\rm M}_{ij}$ (MOSA) about the $x$, $y$, and $z$ axes, respectively, with counterclockwise rotation defined as positive. 
We define the angular jitters as the rotations of incident lasers relative to the MOSAs to stay consistent with the measurements of DWS, which derives these  angles by comparing the wavefronts  of the remote SC's laser beam with that of the local laser beam.
When the jitters are sufficiently small, using a first-order linear approximation, we can write 
\begin{eqnarray}\label{eq:2}
    \phi_{ij} = \hat{x}_{\rm L_{ij}} \cdot \hat{y}_{\rm M_{ij}}, \quad 
    \eta_{ij} = -\hat{x}_{\rm L_{ij}} \cdot \hat{z}_{\rm M_{ij}}.
\end{eqnarray}
Variations in the rolling angle $\theta$ of each MOSA does not introduce considerable  optical path fluctuations~\cite{wanner2024depth}, especially for rotationally symmetric beams,  thus the TTL noise is only associated with the jitters in pitch $\eta$ and yaw $\phi$.

Furthermore, for the measurements of ISIs, the local MOSA receives laser from the remote MOSA. 
We consider the TTL coupling noise in each ISI as the combined effect of received beam on the local MOSA (Rx) and the transmitted beam from the remote MOSA (Tx), and the TTL noise induced by the jitters of remote MOSA needs to be delayed according to the light travel time along this link.  
For any signal or noise $A(t)$, we define the delay operator $D_{ij}$ as 
\begin{equation}
    D_{ij}A(t) \equiv A\left[t-\tau_{ij}(t)\right] .
\end{equation}
Here, $\tau_{ij}(t)$ represents the time it takes for the laser light to travel from SC$_j$ to SC$_i$, i.e.  $\tau_{ij}(t) \equiv | \vec{L} _{ij}(t)|/c$. 
When the jitter angle is sufficiently small, we can approximate that the induced optical path variation is proportional to it by some coupling coefficient $C$. 
In summary, we have the TTL model
\begin{eqnarray}
    \xi_{ij}^{\rm TTL} =  C_{ij\eta Rx} \cdot \eta_{ij} + C_{ji\eta Tx} \cdot D_{ij}\eta_{ji} + C_{ij\phi Rx} \cdot \phi_{ij}+C_{ji\phi Tx} \cdot D_{ij}\phi_{ji}.
    \label{TTL model}
\end{eqnarray}
With $i,j\in \{1,2,3\}$, $k \in \{\phi,\eta\}$ and $l \in \{{\rm Rx},{\rm Tx}\}$, we establish all 24 TTL coefficients $C_{ijkl}$. Ref.~\cite{paczkowski2022postprocessing,wanner2024depth} reported that these coefficients are supposed to reach the 10 mm/rad level before the telescope realignment optimization, and reduce to about 2.3 mm/rad after. 
We will utilize these two magnitudes in the subsequent analysis as representative examples for different statuses of the payloads, accounting for the potential imperfections of alignment. 
On the other hand, the pitch $\eta$ and yaw $\phi$ angles can be measured via the DWS technique to  sub-${\rm nrad}$ precision, making it possible to fit and subtract TTL noise in the data processing stage. 
The readouts of DWS can be regarded as the superposition of angular jitters and readout noises.
Above models will serve as the basis for the simulation and subtraction of TTL noise.

\subsection{The second-generation time-delay interferometry}\label{subsec:tdi}
In space-based GW interferometric measurements, due to the orbital motions of  the three SCs, the arm lengths of these inter-spacecraft interferometers may differ up to a percent level~\cite{Wang:2017aqq,PhysRevD.106.102005,Zhang:2024zxj}. 
Given such unequal arm lengths, laser frequency fluctuations will experience different delays along different arms. 
This can lead to laser frequency noises that are of 6 - 8 orders of magnitude larger than the expected GW signals. 
To resolve such challenge, the TDI method was proposed to synthesize equal-arm interferometry out of the one-way measurements, hence effectively suppressing the laser frequency noises~\cite{tinto1999cancellation,armstrong1999time,tinto2021time}. 
Taking the interferometers on SC$_1$ as an example, the original laser interferometric readout can be expressed as:
\begin{eqnarray}
    s_{12} &=& H_{12}+D_{\rm 12}p_{\rm 21} -p_{12}+D_{\rm 12}k_{21}(\vec{n}_{21}\cdot D_{\rm 12}\vec{\Delta}_{21} + \vec{n}_{12} \cdot \vec{\Delta}_{12}+\xi_{12}^{\rm TTL}), \nonumber\\ 
    \varepsilon_{12} &=& p_{12} - p_{13} - \mu_{13} - k_{12}(2 \vec{n}_{12} \cdot \vec{\Delta}_{12}), \nonumber\\
    \tau_{12} &=& p_{12} - p_{13} - \mu_{13}, \nonumber\\
    s_{13} &=& H_{13}+D_{\rm 13}p_{31} -p_{13}+D_{\rm 13}k_{31}(\vec{n}_{31}\cdot D_{\rm 13}\vec{\Delta}_{31} +\vec{n}_{13} \cdot \vec{\Delta}_{13}+\xi_{13}^{\rm TTL}), \nonumber\\
    \varepsilon_{13} &=& p_{13} - p_{12} - \mu_{12} - k_{13}(2 \vec{n}_{13} \cdot \vec{\Delta}_{13}), \nonumber\\
    \tau_{13} &=& p_{13} - p_{12} - \mu_{12},
\end{eqnarray}
where $s$ represents the readout from the ISI, $\varepsilon$ represents the readout from the TMI, and $\tau$ represents the readout from the reference interferometer (RFI). We use $H$ to represent the GW signal, $p$ the laser frequency noise, and $\mu$ the fiber backward link noise, with all of these quantities defined in terms of phase. $\vec{\Delta}$ represents the displacement vector of OB. $\vec{n}$ denotes the unit vector in the direction of the corresponding arm $\vec{L}$. Therefore, it is necessary to multiply by the corresponding wavenumber $k$ to convert the optical path variations or displacement terms into phase data.

 TDI processing can be performed in three steps~\cite{initial_clock_noise,Otto_thesis,Hartwig_thesis}. 
At first, we construct the intermediate variables $\overline{\xi}$:
\begin{equation}
    \overline{\xi}_{ij}=s_{ij}+\frac{D_{ij}k_{ji}}{k_{ij}}\frac{\varepsilon_{ij}-\tau_{ij}}{2}+D_{ij}\frac{\varepsilon_{ij}-\tau_{ij}}{2}, 
\end{equation}
which is free from noise caused by OB displacements $\vec{\Delta}$. Next, we construct another intermediate variable $\overline{\eta}$:
\begin{eqnarray}
\overline{\eta}_{ij} = \overline{\xi}_{ij}+D_{ij}\frac{\tau_{jk}-\tau_{ji}}{2}, \quad 
\overline{\eta}_{ik} = \overline{\xi}_{ik}+\frac{\tau_{ik}-\tau_{ij}}{2}.\end{eqnarray}
Note that when only TTL noise is present, 
we have $\overline{\eta}_{ij}^{\rm TTL} = \xi_{ij}^{\rm TTL}$.

When the rate of change of arm lengths over time is sufficiently large, the first-generation TDI technology will no longer suffice, as the residual laser frequency noise remains above the required noise allocation threshold. 
The second-generation TDI considers the first-order linear approximate variation of arm length with time and can meet the requirements for space-based GW detection~\cite{Tinto:2022zmf}. 
The output of the second-generation Michelson $X$ channel is 
\begin{eqnarray}
    X_2 =& (1-D_{12}D_{21}-D_{12}D_{21}D_{13}D_{31} \nonumber\\
    &+D_{13}D_{31}D_{12}D_{21}D_{12}D_{21})  (\overline{\eta}_{13}+D_{13}\overline{\eta}_{31}) \nonumber\\ 
    &- (1-D_{13}D_{31}-D_{13}D_{31}D_{12}D_{21} \nonumber\\&+D_{12}D_{21}D_{13}D_{31}D_{13}D_{31})  (\overline{\eta}_{12}+D_{12}\overline{\eta}_{21}),
    \label{TDI}
\end{eqnarray}
where  the laser frequency noise is successfully suppressed. 
Cycling the subscripts with rule ``1 $\rightarrow$ 2 $\rightarrow$ 3 $\rightarrow$ 1'', we can obtain two more TDI combinations $Y_2$ and $Z_2$. 
By employing the linear TTL model and the rules of TDI combination, we obtain the final TTL model within the TDI data, which can be further utilized to construct the likelihood function.

Since $\overline{\eta}_{ij}^{\rm TTL} = \xi_{ij}^{\rm TTL}$ when only TTL noise is present, we can directly substitute  Eq.~(\ref{TTL model})  into  Eq.~(\ref{TDI}), obtaining the total TTL noise in the $X_2$ variable. 
For the $X_2$ channel, we observe that the contributions  from 16 TTL coefficients are non-zero, as the jitter of MOSA$_{23}$ and MOSA$_{32}$ have no influence on optical paths of $X_2$ channel.
To estimate all the 24 TTL coefficients, we  need to
jointly analyze the $X_2, Y_2, Z_2$ channels.

Due to the linear nature of our TTL model, 
we can model TTL noise in the TDI channels as the sum of contributions from 24 TTL coefficients.
The contribution of each coefficient  can be isolated  by setting the other 23 coefficients to 0. 
For example, according to Eq.~(\ref{TTL model}) and Eq.~(\ref{TDI}), the contribution from $C_{12\eta Rx}$ to the $X_2$ channel is 
\begin{eqnarray}
    C_{12\eta Rx} X_{C_{12\eta Rx}}^{\rm TTL}  &\equiv& - (1-D_{13}D_{31}-D_{13}D_{31}D_{12}D_{21} \nonumber\\
    && +D_{12}D_{21}D_{13}D_{31}D_{13}D_{31}) \cdot C_{12\eta Rx}\cdot \eta_{12}.
    \label{TTL贡献}
\end{eqnarray}
Therefore the total TTL noise in the  $X_2$ channel reads 
\begin{equation}
    X_2^{\rm TTL} = \sum_{ijkl} C_{ijkl} X^{\rm TTL}_{2 \ C_{ijkl}}.
\end{equation}

\subsection{Drag-free attitude control system}
As is discussed in Section~\ref{sec:I}, the angular jitters of the SCs and OBs are essentially the residual noise of drag-free control. 
Therefore, in principle a practical TTL noise subtraction algorithm should be developed and tested based on the simulation of DFACS.
In this subsection, we provide a concise overview on the objectives and technical methodologies used by DFACS.

DFACS, ensuring translation and attitude of SCs, TMs and MOSAs to stay stably at their nominal position, is designed to fulfill the following three objectives: 
\begin{enumerate}
    \item The SC translation is forced to follow the average trajectories of two TMs, which approximately move along geodesics. According to local TM interferometry measurement, proper actuation from the micro-thrusters is applied on the SC to compensate stray and external forces, such as solar pressure, celestial gravity gradient, and so on, to make sure the spacecraft to be nearly as quite as the testing masses at their $x$-axis directions. (the relative displacement offset $< \sim$10 nm/$\sqrt{\rm Hz}$).
    \item The attitudes of SCs and MOSAs are commanded to make sure that both MOSAs pointing direction track the incoming laser beam direction and nulls DWS output, which guarantees the inter-satellite laser interferometric measurement. The residual pointing jitter is kept below 10 - 30 nrad/$\sqrt{\rm Hz}$ at Taiji band on account of TTL coupling noise mitigation.
    \item To the exclusion of drag-free axes, any relative displacement offset of TMs relative to the corresponding SCs, mainly resulting from SC self-gravity gradients and attitude movement, are corrected by electrostatic forces and torques generated by electrostatic actuation of the gravitational reference system. Because of the cross-talk coupling effect between different degrees of freedom of the electrostatic actuation, a minimal electrostatic interaction is expected to mitigate residual acceleration pollution on drag-free axes. 
\end{enumerate}

Taiji is a complex system involving multibody dynamics, internal and external noise, multiple measurement and actuation systems, and multiloop controller. To characterize the TTL noise caused by attitude jitter of the MOSAs and the SCs as real as possible, the closed-loop simulation is constructed based on nonlinear dynamical model and discrete DFACS. Furthermore, extended precision numerical calculation is employed to avoid extra noise from round-off error.

\begin{figure}[ht]
\centering
\includegraphics[width=0.49\textwidth]{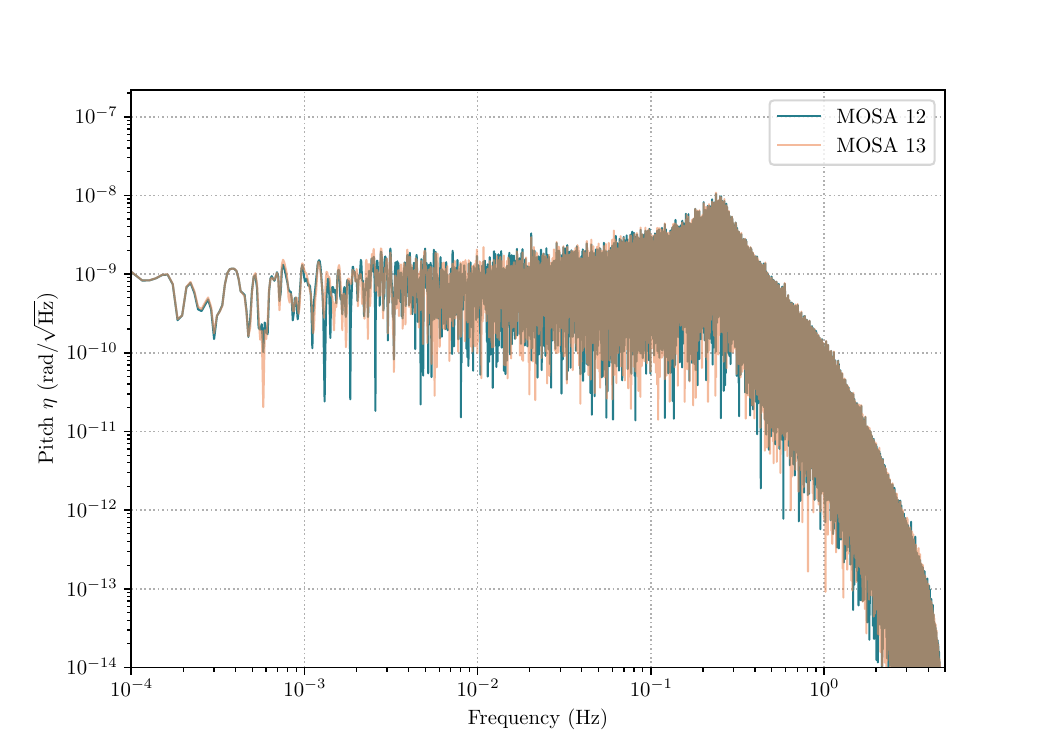}
\includegraphics[width=0.49\textwidth]{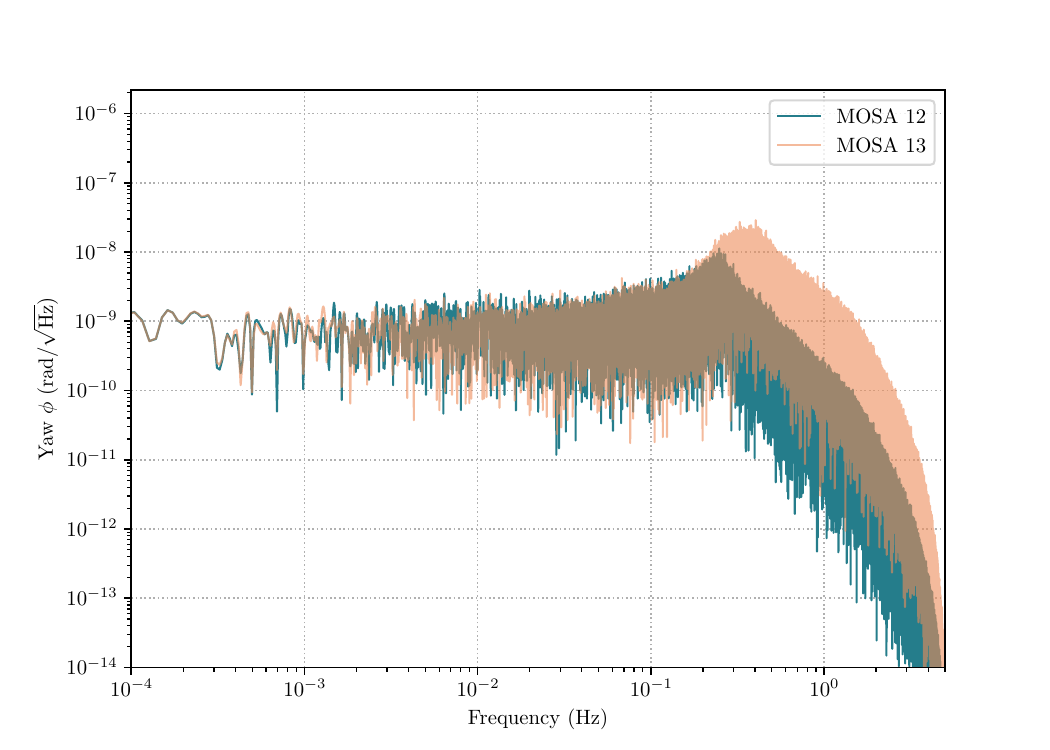}
\caption{The ASDs of SC$_1$'s angular jitters during the course of one day. The left panel represents the jitters of pitches $\eta$, while the right plot shows the jitters of yaws $\phi$.}
\label{TTL仿真1} 
\end{figure}

The design and optimization  of  Taiji's DFACS have been systematically investigated in our forthcoming paper~\cite{qianjiao2024design}, and the DFACS simulation therein serves as a fundamental input of this study.
The  simulation of Ref.~\cite{qianjiao2024design} outputs the rotation matrix of each MOSA$_{ij}$ relative to the corresponding SC$_i$ (dubbed $R^{{\rm SC}_i}_{{\rm M}_{ij}}$), as well as the rotation matrix of each SC$_i$ relative to the inertial frame (IF) (dubbed $R^{\rm{IF}}_{{\rm SC}_i}$). 
Denoting the coordinate components of $\hat{x}_{{\rm M}_{ij}}$  relative to the ${\rm M}_{ij}$ frame as $\hat{x}_{{\rm M}_{ij}}^{{\rm M}_{ij}} = [1, 0, 0]$, then its coordinate components relative to the IF are
\begin{equation}\label{eq:1}
    \hat{x}_{\rm M_{ij}}=R^{\rm{IF}}_{{\rm SC}_i} R^{{\rm SC}_i}_{{\rm M}_{ij}} \hat{x}_{{\rm M}_{ij}}^{{\rm M}_{ij}}.
\end{equation}
Using the SCs' positions at the emission time and reception time of lasers obtained through DFACS simulation, along with the definitions in Section~\ref{subsec:ttl_model}, the components of $\hat{x}_{{\rm L}_{ij}}, \hat{y}_{{\rm M}_{ij}}, \hat{z}_{{\rm M}_{ij}}$ in IF can be calculated.
According to  Eq.~(\ref{eq:2}), 
we further deduce the angular jitters $\phi_{ij}$ and $\eta_{ij}$.
Shown in Figure~\ref{TTL仿真1} are the amplitude spectral densities (ASDs) of angular jitters over the course of one day. 
As is described in Section~\ref{subsec:ttl_model}, the left MOSAs are fixed to SCs while the right MOSAs can rotate around the $z$-axes of SCs, 
which subsequently results in slightly larger jitters of $\phi$ on the right side.
It is demonstrated in Ref.~\cite{qianjiao2024design} that such a design can fulfill the objectives of DFACS.

\subsection{Simulation}
\begin{figure}
\centering
\includegraphics[width=1\textwidth]{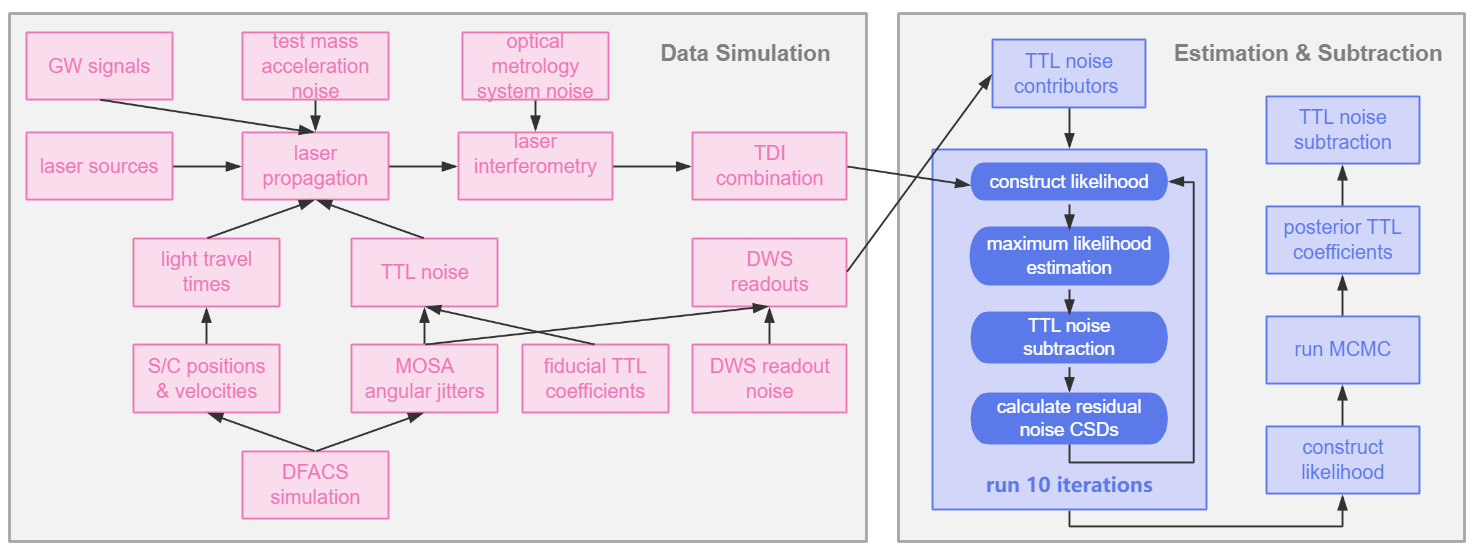}
\caption{The flowchart of data simulation and TTL coefficient estimation.}
\label{系数估计流程} 
\end{figure}

The workflow for data simulation and TTL noise reduction is presented Figure~\ref{系数估计流程}, with the key  simulation  setups summarized  as follows.  
As  explained in Section~\ref{sec:II} and Ref.~\cite{wanner2024depth}, we ignore the TTL noises of TMIs and set them to zero. 
We randomly generate 24 TTL coefficients with a standard deviation of 2.3 mm/rad (or 10 mm/rad), and calculate the TTL noises of ISIs according to Eq.~(\ref{TTL model}). 
Additionally, we have also incorporated the
ACC noises of TMs, the OMS noises of interferometers, as well as the readout noises of DWS, the former two being the main contributors to the noise floors of LISA and Taiji. 
These noises are simulated as Gaussian stationary noises according to their PSDs: 
\begin{eqnarray}\label{eq:2ndnoise}
    P_{\rm OMS}(f) &=& A_{\rm OMS}^2 \left[1 + \left(\frac{2 \ {\rm mHz}}{f}\right)^4\right],  \nonumber \\
    P_{\rm ACC}(f) &=& A_{\rm ACC}^2   \left[1 + \left(\frac{0.4 \ {\rm mHz}}{f}\right)^2\right] \left[1 + \left(\frac{f}{8 \ {\rm mHz}}\right)^4\right], \nonumber \\ 
    P_{\rm DWS}(f) &=& A_{\rm DWS}^2 \left[1 + \left(\frac{2 \ {\rm mHz}}{f}\right)^4\right],
\end{eqnarray}
where the noise amplitudes are  $A_{\rm OMS}=8 \ {\rm pm/\sqrt{Hz}}$, $A_{\rm ACC} = 3 \ {\rm fm / s^2 / \sqrt{Hz}}$, and $A_{\rm DWS}=0.21 \ {\rm nrad/\sqrt{Hz}}$ respectively~\cite{luo2021taiji}. 
Besides, we assume all the clocks on-board the SCs are perfectly synchronized.
The duration of data is set to 1 day, simulated at the  sampling rate of 10 Hz.  
To demonstrate the robustness of our algorithm in the presence of GWs, a GW signal from MBHB  is injected into the data.
The  source  has a (redshifted) chirp mass of $5\times 10^4 M_{\bigodot}$ and mass ratio 0.36. 
It is located at redshift 1, corresponding to a luminosity distance of 6791 Mpc,  with randomly generated Ecliptic longitude and latitude of 3.03 rad and 0.38 rad, respectively. 
The merger time is placed at the 60,000 s of the day. 

\begin{figure}
\centering
\includegraphics[width=0.6\textwidth]{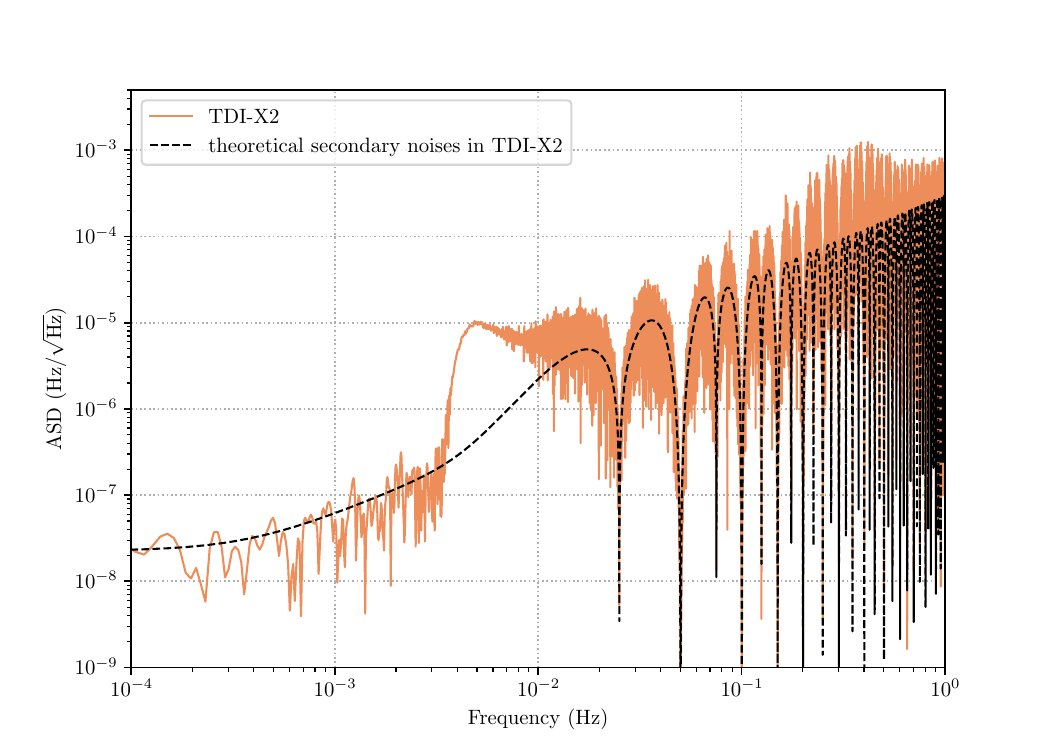}
\caption{The orange curve shows the ASD of simulated $X_2$ data stream with $C_{ijkl} \sim $ 2.3 mm/rad, and the theoretical ASD of secondary noises in the $X_2$ channel is shown as black dashed curve. 
As can be seen, TTL noise excesses the secondary noises especially  at the 0.1 Hz - 1 Hz band, by half to one order of magnitude. 
The bump at 1 $\sim$ 10 mHz is caused by the injected GW signal.}
\label{X2输出} 
\end{figure}

Figure~\ref{X2输出} shows the simulated TDI-$X_2$ data stream with TTL coefficients at the 2.3 mm/rad level (also see the red curves in Figure~\ref{2.3结果} for better visualization of the GW signal). 
From Figure~\ref{X2输出}, we can see that the TTL noise is half to one order of magnitude larger than the total secondary noise budget of Taiji, hence subsequent suppression of TTL noise is indispensable.

\section{TTL noise subtraction algorithm\label{sec:III}}
\subsection{Bayesian estimation of TTL coefficients}
In this section, we introduce our enhanced  algorithm for TTL noise suppression. 
The model of TTL noise in each TDI variable is established in Section~\ref{subsec:tdi}. 
By integrating the  TDI data, the parametric TTL model,  and the covariance matrix between TDI channels, we formulate  the  matched filtering likelihood function as follows:
\begin{eqnarray}
    {\rm ln}\mathcal{L}(\boldsymbol{C}) = -\frac{1}{2} \left(\tilde{\boldsymbol{d}} - \tilde{\boldsymbol{h}}(\boldsymbol{C})\right)^\dagger \boldsymbol{Cov}^{-1} \left(\tilde{\boldsymbol{d}} - \tilde{\boldsymbol{h}}(\boldsymbol{C})\right),
    \label{likelihood function}
\end{eqnarray}
where $\tilde{\boldsymbol{d}}\equiv[\tilde{X_2},\tilde{Y_2}, \tilde{Z_2}]$ represents  the data of the $X_2$, $Y_2$ and $Z_2$ channels, and $\tilde{\boldsymbol{h}}\equiv [\tilde X^{\rm TTL}_2,\tilde Y^{\rm TTL}_2,\tilde Z^{\rm TTL}_2]$ denote the modeled TTL noise, which  depends on the parameter set $\boldsymbol{C} \equiv  \{C_{ijkl}\}$. 
All the angular jitter terms in $\tilde{\boldsymbol{h}}(\boldsymbol{C})$ should be replaced by the corresponding DWS readouts. 
The tilde symbol ``$\sim$'' indicates taking the Fourier transform.
The covariance matrix $\boldsymbol{Cov}$ reads
\begin{eqnarray}
    \boldsymbol{Cov} = 
        \left[ 
        \begin{array}{ccc}  
        \boldsymbol{C}_{X_2X_2^*} & \boldsymbol{C}_{X_2Y_2^*} & \boldsymbol{C}_{X_2Z_2^*} \\  
        \boldsymbol{C}_{Y_2X_2^*} & \boldsymbol{C}_{Y_2Y_2^*} & \boldsymbol{C}_{Y_2Z_2^*} \\  
        \boldsymbol{C}_{Z_2X_2^*} & \boldsymbol{C}_{Z_2Y_2^*} & \boldsymbol{C}_{Z_2Z_2^*}  
        \end{array}  
        \right]. 
\end{eqnarray}
For example, 
\begin{equation}
    \boldsymbol{C}_{X_2Y_2^*}  = \frac{1}{4\Delta f}\boldsymbol{S}_{X_2Y_2^*},
\end{equation}
where 
$\boldsymbol{S}_{X_2Y_2^*}$ is the Fourier series of channel $X_2$, $Y_2$'s correlation spectral density (CSD). 
Since TTL noise is effectively treated as a ``signal'' in Eq.~(\ref{likelihood function}), $\boldsymbol{Cov}$ actually  represents the covariance of the noise floor. 
However, the amplitude of TTL noise is predicted to be greater than that of the noise floor, thus $\boldsymbol{Cov}$ cannot be determined prior to the subtraction of TTL noise.
This limitation poses a significant challenge, which  our algorithm is specifically designed to address.

According to the principle of matched filtering, optimal parameter estimation requires whitening the residual $\tilde{d} - \tilde{h}$  with the CSD of noise floor. 
However, in practice, we typically do not  know the CSD of the secondary noise beforehand. 
As an initial approximation, we assume the secondary noise to be white, namely the PSDs are all constants (say PSD $\equiv$ 1), and there are no correlations between different TDI channels. 
At this point, the likelihood function reduces to the least-square optimization loss function previously employed in  Ref.~\cite{paczkowski2022postprocessing}. 
Based on this approximation, by maximizing the likelihood function (in this work  the differential evolution algorithm~\cite{DifferentialEvolution} is employed), we can roughly estimate the 24 TTL coefficients.
Subtracting the reconstructed TTL noise from the TDI data, 
we then get a residual noise  and calculate its CSD using the Welch method. 
Through this preliminary optimization and subtraction, we essentially separate the noise floor (which is uncorrelated to the angular jitters) from the TTL noise (which is correlated to the angular jitters). 
This calculated CSD is used to establish a new likelihood function and again estimate the TTL coefficients. 
By performing the aforementioned steps iteratively, the  maximum likelihood estimation (MLE) for the TTL coefficients and the evaluation of noise floor CSD will be refined.

To further determine the TTL coefficients more accurately and obtain their confidence intervals, above MLE iterations should be followed by a  MCMC (Markov Chain Monte Carlo) parameter estimation. 
For this final MCMC estimation we use the multi-purpose sampler \texttt{Eryn}~\cite{Karnesis:2023ras,michael_katz_2023_7705496}  with an \texttt{emcee} move~\cite{2013PASP..125..306F}. 

The complete procedure of TTL noise suppression can be summarized as follows. 
We first calculate the TTL noise contribution using the DWS data, and build the likelihood function according to Eq.~(\ref{likelihood function}). 
After several iterations of MLE, we obtained a more accurate estimate of the noise floor CSD, and used it to construct the final likelihood. 
Finally, an MCMC estimation is performed, yielding not only the 24 TTL coefficients, but also their confidence intervals and  the covariances between each two of them.

The strategy for handling GW signals encompasses two key aspects.
On the one hand, the target frequency band for Taiji is  0.1 mHz - 0.1 Hz.
The presence of GW signals in this band would disrupt the Gaussianity and stationarity of data,  violating the prerequisites for the aforementioned Bayesian formalism.
To mitigate the potential bias caused by  GW signals, we confine TTL coefficient estimation to the  0.05 Hz - 0.5 Hz range. 
Data at $>0.5$ Hz is also excluded because 
practical issues such as the effects of filters and interpolation error might become evident at higher frequencies, as noted by  Refs.~\cite{PhysRevD.99.084023,PhysRevD.109.043040}.
On the other hand, for the few GW signals in the 0.05 Hz - 0.5 Hz frequency band (such as the mergers of low-mass MBHBs), we also treat them as part of the noise floor and process them iteratively alongside other noises.
Although they are still non-Gaussian and non-stationary in nature, the impacts should be negligible due to the relatively small amplitudes.

\subsection{Theoretical analysis with the Fisher formalism}

In order to provide a theoretical estimate for the fitting accuracy for the TTL coefficients, 
hence guiding the design of our  algorithm (such as the duration of data required for sufficient TTL noise suppression), 
we utilize Fisher information matrix (FIM) to evaluate the uncertainties of  coefficients~\cite{vallisneri2008use}. 


For  an order-of-magnitude analysis, we take the single Michelson $X_2$ channel as an example. 
The logarithmic likelihood function reads
\begin{eqnarray}
    & \quad \  {\rm ln}\mathcal{L}(\boldsymbol{C}) \nonumber \\ 
    &= -2 \int_{f_{\rm min}}^{f_{\rm max}} df \frac{|\tilde{X}_2(f) - \tilde{X}_2^{\rm TTL}(f; \boldsymbol{C})|^2}{P^{\rm 2nd}_{X_2}(f)} \nonumber \\
    \nonumber&= -2 \int_{f_{\rm min}}^{f_{\rm max}} df \frac{|\tilde{X}^{\rm 2nd}_2(f) + \tilde{X}_2^{\rm TTL}(f; \boldsymbol{C}_0) - \tilde{X}_2^{\rm TTL}(f; \boldsymbol{C})|^2}{P^{\rm 2nd}_{X_2}(f)} \\
    \nonumber&= -2 \int_{f_{\rm min}}^{f_{\rm max}} df \frac{|\tilde{X}^{\rm 2nd}_2(f) - \sum_{ijkl} \tilde{X}^{\rm TTL}_{2 \ C_{ijkl}}(f) \Delta C_{ijkl}|^2}{P^{\rm 2nd}_{X_2}(f)} \\
    \nonumber&\approx -\sum_{f_a}  \left[ 1 + \frac{2\Delta f |\sum_{ijkl} \tilde{X}^{\rm TTL}_{2 \ C_{ijkl}}(f_a) \Delta C_{ijkl}|^2}{P^{\rm 2nd}_{X_2}(f_a)} \right].
\end{eqnarray}
The third line results from the fact that $\tilde{X}_2(f)$ is the combination of secondary noises $\tilde{X}^{\rm 2nd}_2(f)$ and TTL noise $\tilde{X}_2^{\rm TTL}(f; \boldsymbol{C}_0)$. In the forth line, the deviation of $\boldsymbol{C}$ from the ``true'' coefficients $\boldsymbol{C}_0$ is denoted as  $\Delta C_{ijkl}$, and $\tilde{X}^{\rm TTL}_{2 \ C_{ijkl}}(f)$ is the TTL contribution defined in Eq.~(\ref{TTL贡献}).  
To deduce the fifth line we utilized the statistical independence between TTL noise and secondary noises. 
By definition, the diagonal elements of the FIM are
\begin{eqnarray}
    \Gamma_{C_{ijkl}C_{ijkl}} &=& -\bigg\langle \frac{\partial^2 {\rm ln}\mathcal{L}(\boldsymbol{C})}{\partial C_{ijkl}^2} \bigg\rangle \approx \sum_{f_a}  \bigg\langle \frac{4\Delta f  |\tilde{X}^{\rm TTL}_{2 \ C_{ijkl}}(f_a)|^2 }{P^{\rm 2nd}_{X_2}(f_a)} \bigg\rangle \nonumber  \\
    &\approx& \sum_{f_a}  \frac{2 P_{X_2}^{C_{ijkl}}(f_a) }{P^{\rm 2nd}_{X_2}(f_a)},
\end{eqnarray}
where $P_{X_2}^{C_{ijkl}}$ is the PSD of $\tilde{X}^{\rm TTL}_{2 \ C_{ijkl}}$. 
For $ijkl = 12\eta{\rm Rx}$, according to Eq.~(\ref{TTL贡献}), one has 
\begin{equation}
    \Gamma_{C_{12\eta{\rm Rx}}C_{12\eta{\rm Rx}}} \approx 2\sum_{f_a}  \frac{ P_{X_2}^{C_{12\eta{\rm Rx}}}(f_a) }{P^{\rm 2nd}_{X_2}(f_a)} \approx  \frac{N_f}{2} \left(\frac{A_\eta}{A_{\rm OMS}}\right)^2,
\end{equation}
$N_f$ being the number of frequency-domain data points. 
In the above derivation we have neglected the acceleration noise since the likelihood is calculated in $f \in [f_{\rm min}, f_{\rm max}] = [0.05, 0.5]\ {\rm Hz}$ where OMS noise dominates.
The readout noise of DWS is also neglected due to its relatively small amplitude. 
In this order-of-magnitude analysis  we assume that OMS noises and angular jitters are both white noises with amplitudes $A_{\rm OMS}$ and $A_\eta$, respectively. 
Assuming that the coefficients are weakly correlated, 
it follows that the lower bound of $C_{12\eta{\rm Rx}}$'s $1\sigma$ uncertainty is 
\begin{eqnarray}\label{eq:confidence_interval}
    \sigma_{C_{12\eta{\rm Rx}}} &\ge \left(\Gamma_{C_{12\eta{\rm Rx}}C_{12\eta{\rm Rx}}}\right)^{-1/2} 
    \approx \frac{A_{\rm OMS}}{A_\eta} \sqrt{\frac{2}{N_f}} \nonumber \\ 
    & = \frac{A_{\rm OMS}}{A_\eta} \sqrt{\frac{2 }{\left(f_{\rm max} - f_{\rm min}\right) T}}, 
\end{eqnarray}
according to the interpretation of FIM. 
Consistent with our expectations, $\sigma_{C_{12\eta{\rm Rx}}}$ is inversely proportional to  the square root of data duration $T$. 
For instance, if $T = $ 1 day, $A_{\rm OMS}=8 \ {\rm pm/\sqrt{Hz}}$ and $A_{\eta} = 1 \ {\rm nrad/\sqrt{Hz}}$, then  $\sigma_{C_{12\eta{\rm Rx}}} \geq 0.057  \ {\rm mm/rad}$. 
This level of uncertainty corresponds to a residual TTL noise far below the amplitudes of secondary noises. 
Besides, since each contribution $\tilde{X}^{\rm TTL}_{2 \ C_{ijkl}}$ is proportional to the amplitude of corresponding angular jitter $A_k$ (with $k = \eta$ or $\varphi$), after multiplied by $\sigma_{C_{ijkl}}$ and summing over $ijkl$,
the resulting residual TTL noise should be theoretically independent of $A_k$'s amplitudes. 
Moreover, our preliminary analysis indicates that the dependence of residual TTL noise on the magnitudes of $C_{ijkl}$ is also not significant, as $C_{ijkl}$ are absent from Eq.~(\ref{eq:confidence_interval}).

\section{Results\label{sec:IV}}

\begin{figure}
    \centering
    \includegraphics[width=0.6\textwidth]{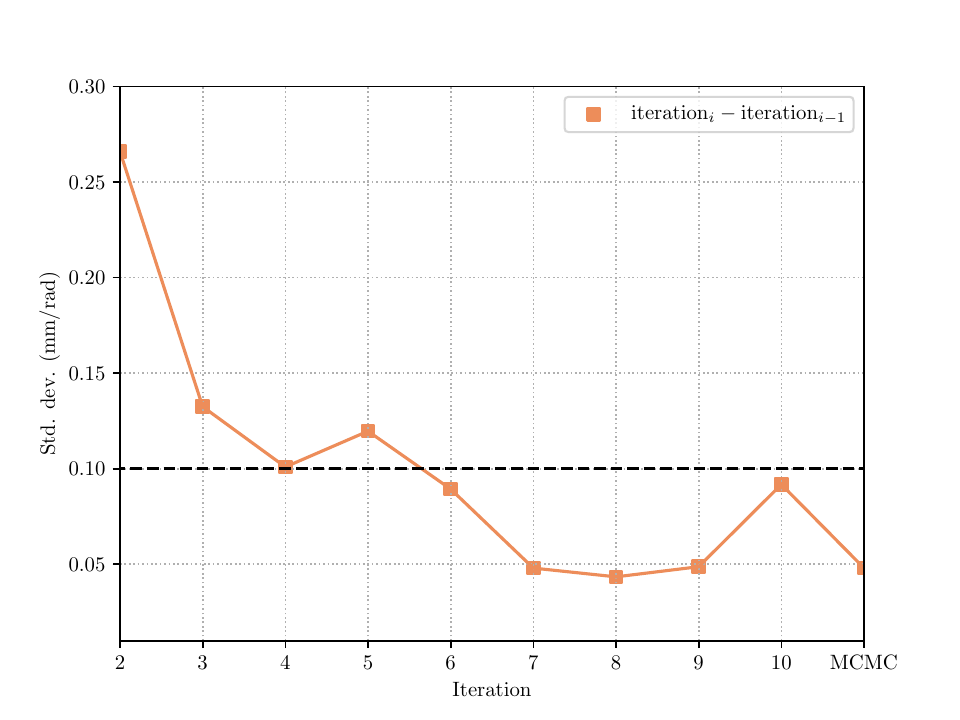}
    \caption{The iteration process. The orange line represents the standard deviations of the coefficients obtained in each iteration compared to the  previous iteration, and the black dashed line labels the 0.1 mm/rad threshold.} 
    \label{迭代过程} 
\end{figure}

\begin{table}
    \centering
    \caption{The true values of TTL coefficients, results and deviations after the 1st iteration, results and deviations after the final MCMC estimation, unit: mm/rad.\label{tab:I}} 
    \begin{tabular}{cccccc}
    \toprule
    \textbf{parameter} & \textbf{truth} & \textbf{1st coef.} & \textbf{1st dev.} & \textbf{final coef.} & \textbf{final dev.}  \\
    \midrule
    12$\phi$Tx & 0.517 & 0.599 & 0.082 & 0.531 & 0.014 \\
    13$\phi$Tx & -1.233 & -1.154 & 0.079 & -1.207 & 0.026  \\
    23$\phi$Tx & 1.180 & 1.213 & 0.033 & 1.179 & -0.001  \\
    21$\phi$Tx & -2.117 & -2.033 & 0.083 & -2.161 & -0.044  \\
    31$\phi$Tx & 4.538 & 4.578 & 0.041 & 4.550 & 0.012 \\
    32$\phi$Tx & 1.300 & 1.372 & 0.072 & 1.261 & -0.039 \\
    12$\eta$Tx & -1.824 & -3.987 & -2.163 & -2.081 & -0.257  \\
    13$\eta$Tx & 0.188 & -1.866 & -2.054 & 0.010 & -0.177  \\
    23$\eta$Tx & -2.259 & -1.790 & 0.469 & -2.140 & 0.119 \\
    21$\eta$Tx & -0.802 & -0.069 & 0.733 & -0.654 & 0.147 \\
    31$\eta$Tx & -0.849 & -0.135 & 0.713 & -1.106 & -0.257 \\
    32$\eta$Tx & -0.613 & -0.117 & 0.496 & -0.828 & -0.215 \\
    12$\phi$Rx & -1.854 & -1.877 & -0.023 & -1.863 & -0.008  \\
    13$\phi$Rx & -0.274 & -0.471 & -0.197 & -0.308 & -0.034  \\
    23$\phi$Rx & -1.057 & -1.093 & -0.037 & -1.052 & 0.005  \\
    21$\phi$Rx & -3.633 & -3.747 & -0.114 & -3.630 & 0.003  \\
    31$\phi$Rx & 2.480 & 2.478 & -0.002 & 2.462 & -0.018  \\
    32$\phi$Rx & -1.239 & -1.116 & 0.123 & -1.140 & 0.099  \\
    12$\eta$Rx & 1.406 & 3.497 & 2.091 & 1.593 & 0.187 \\
    13$\eta$Rx & -3.756 & -1.507 & 2.249 & -3.600 & 0.156 \\
    23$\eta$Rx & -1.599 & -2.520 & -0.922 & -1.711 & -0.112  \\
    21$\eta$Rx & -3.271 & -4.023 & -0.752 & -3.304 & -0.033  \\
    31$\eta$Rx & -0.633 & -1.189 & -0.557 & -0.419 & 0.214  \\
    32$\eta$Rx & 0.667 & 0.154 & -0.513 & 0.848 & 0.181  \\
    \bottomrule
    \end{tabular}   
\end{table}

This section presents the  estimation results based on numerical simulation.
In the initial MLE iteration, the PSD  used in the likelihood function is set to 1. 
Following the first iteration, we evaluate the CSD after TTL noise subtraction,  and substituting it back into the likelihood function.
By maximizing  the updated likelihood,   a refined  MLE of the TTL coefficients is obtained.  
Theoretically, these coefficients will be closer to the true values than  the first iteration since we have better knowledge on  the noise floor. 
Through iterative refinement, the MLEs of  TTL coefficients are expected to  converge towards  a stable set of  values.
For the case where the  TTL coefficients are set to  2.3 mm/rad level,  we performed 10 MLE iterations. 
In realistic  detection,  one does not have access to the ``true'' parameters, thus we assess the convergence of iteration by monitoring the differences between consecutive iterations.
The  iteration process is visualized in Figure~\ref{迭代过程}, 
where the orange line represents the standard deviations of the coefficients obtained in each iteration compared to the previous iteration,
and the black dotted line representing the 0.1 mm/rad threshold.
As expected, the differences between consecutive MLEs exhibit oscillatory decay as the number of iterations increases,  and ultimately  approaches the threshold.

\begin{figure}
\centering
\includegraphics[width=0.7\textwidth]{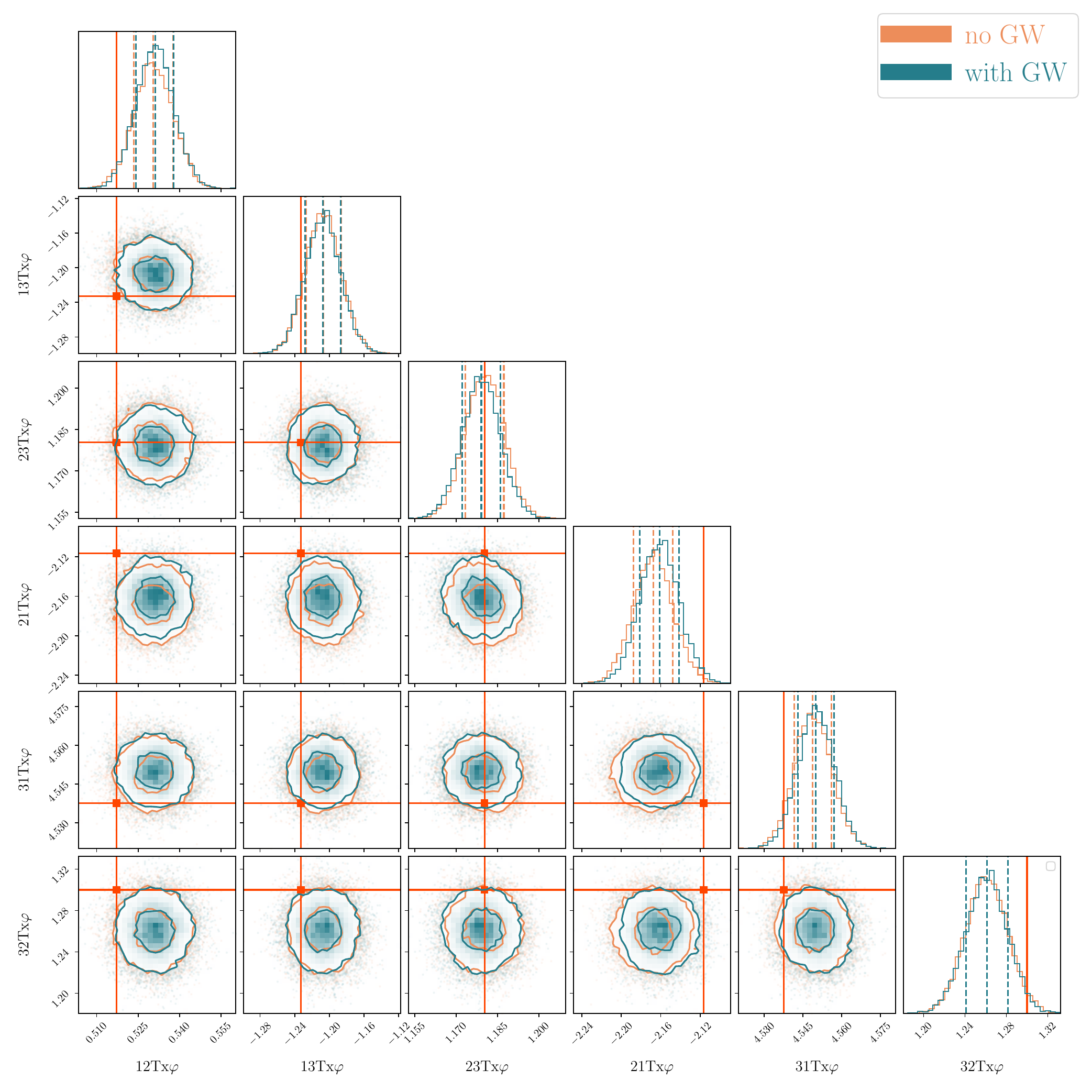}
\caption{Posterior distributions of the first 6 TTL  coefficients. 
The green and orange corner plots represent the results of simulation with / without GW signal, respectively. 
As is shown in the figure, under our algorithm design, the injected GW signal does not have considerable impact on the estimation.}
\label{方差协方差矩阵} 
\end{figure}

The fiducial TTL coefficients used in simulation, alongside the estimated coefficients after the first MLE run and the final MCMC run, are presented in Table~\ref{tab:I}. 
The results of MCMC are significantly better than those of the first MLE, \textcolor{black}{with a standard deviation of $\sim$ 0.1 mm/rad relative to the fiducial values.  }
Through the final MCMC estimation, we also obtain the posterior distributions of all the 24 TTL coefficients. 
For a clear display of the distributions, 
the posteriors of the first 6 coefficients are shown in Figure~\ref{方差协方差矩阵} as green corner plot.
For comparison, we have also presented the result of simulation without GW signal as orange corner plot in the same figure. 
As can be seen, 
the injected $5\times 10^4 M_{\bigodot}$ MBHB signal does not have considerable impact on the results. 

\begin{figure}
\centering
\includegraphics[width=0.495\textwidth]{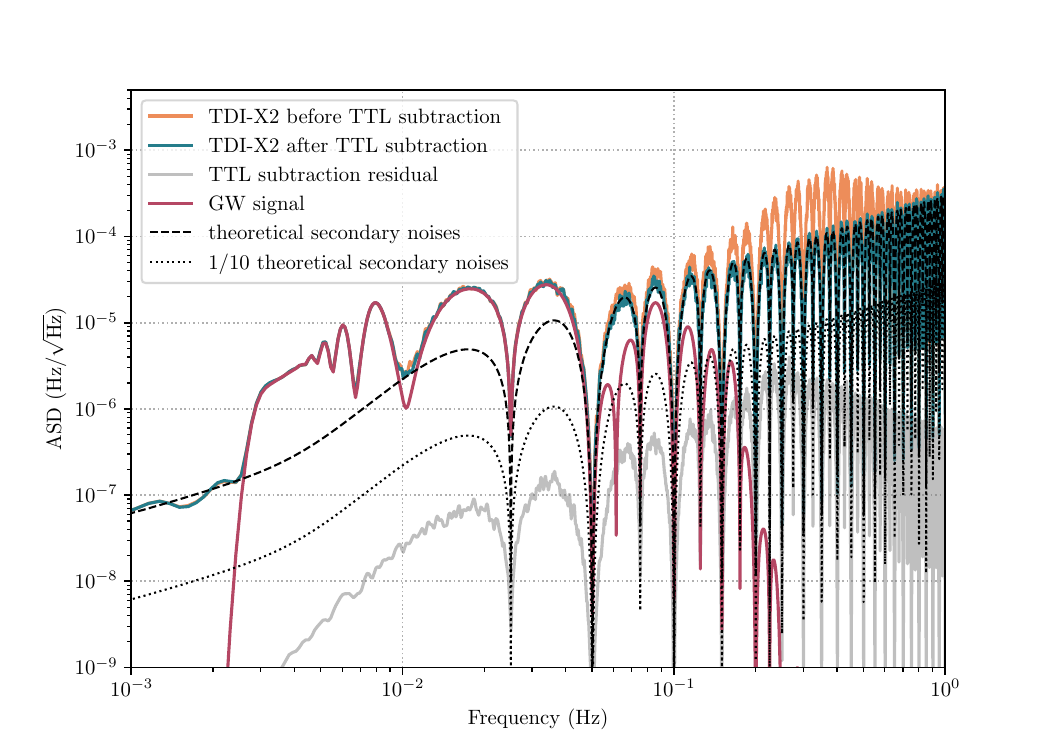}
\includegraphics[width=0.495\textwidth]{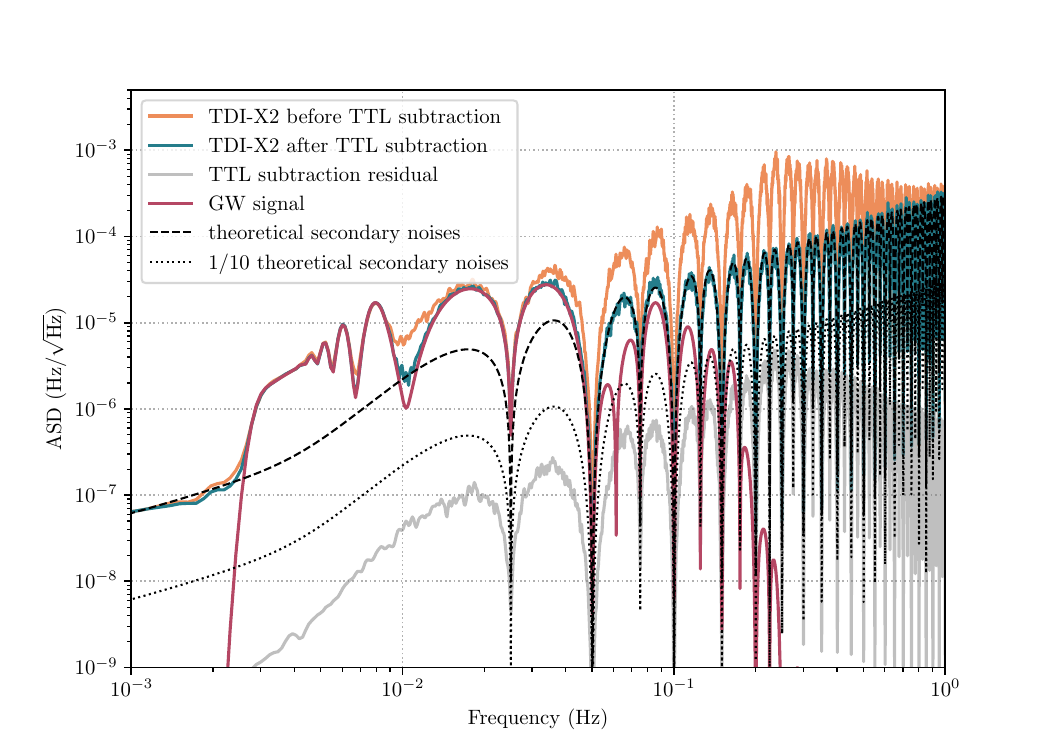}
\caption{The subtraction results of the $X_2$ channel  with  TTL noise coefficients at the 2.3 mm/rad level (left panel) and 10 mm/rad level (right panel).}
\label{2.3结果} 
\end{figure}

The subtraction result for the $X_2$ channel is shown in the left panel of Figure~\ref{2.3结果}. 
The ASDs of TDI-$X_2$ channel before and after TTL noise subtraction are plotted as orange and green curves respectively, and the red curve stands for the injected GW signal. 
The theoretical ASD of the secondary noises, as well as the 1/10 of it, are also shown in the figure, serving as criteria for the accuracy of TTL noise subtraction. 
The residual TTL noise after subtraction is  displayed as a grey curve.
To clearly demonstrate the spectral shapes of the noises and signal, the aforementioned ASDs are calculated using the averaged Welch's method. 
Notably, the residual TTL noise is more than one order of magnitude lower than the secondary noises. 
To account for the situation where the telescope alignment may not be perfect, we also conducted another  simulation by elevating the TTL noise coefficient to the 10 mm/rad level. 
After implementing the same TTL noise subtraction process, we observed results  illustrated in the right panel of Figure~\ref{2.3结果}. 
Still, the residual TTL noise remains more than one order of magnitude lower than the secondary noise,  fulfilling the requirements of both LISA and Taiji.
These results, along with the Fisher analysis in the previous section, indicate that
our algorithm is robust and applicable to different statuses of the payloads.

\section{Conclusions\label{sec:V}}
After suppressing laser frequency noise through TDI, TTL noise is predicted to be one of the main noise sources for the LISA and Taiji missions. 
In this paper, we designed an enhanced algorithm,
aiming to effectively fit and subtract it during the  data processing stage. 
The primary contribution of this algorithm is to  address limitations in  realistic  detection scenarios.

We have re-derived the model of TTL coupling noise  in  the second-generation  Michelson TDI variables, based on the rotation matrices of  SCs and  MOSAs generated from DFACS simulation. 
Besides, our algorithm simultaneously estimates the TTL coefficient and the  cross spectra of noise floor  through an iterative MLE procedure, 
thereby addressing the issue of unknown noise floor during realistic detection. 
The iteration is then followed by an MCMC parameter estimation to refine the TTL coefficients  and quantify their confidence intervals. 
Additionally, we have accounted for the impacts of GW signals. 
To mitigate the potential biases caused by GW signals, we only use data within the frequency range of  0.05 Hz to 0.5 Hz to construct the likelihood function.

Through Fisher information analysis, we theoretically assessed the uncertainty of estimation and found that neither the magnitudes of TTL coefficients nor the angular jitters could considerably affect the result of TTL noise subtraction, thereby demonstrating   our algorithm is  applicable to diverse payload configurations  and robust to potential imperfections. 

Results of numerical simulations reveal that the differences between the estimated coefficients and their true values have a standard deviation of  $\sim$ 0.1 mm/rad. 
For TTL coefficients at both 2.3 and 10 mm/rad levels, the residual TTL noises after subtraction are more than one order of magnitudes below  the secondary noises, fulfilling the mission requirements of LISA and Taiji, hence 
paving the way for future picometer-precision  GW  observation.

\section*{Data Availability}
The simulated data included in this study are available upon request by contact with the corresponding author. 


\ack 
This study is supported by the National Key Research and Development Program of China (Grant No. 2021YFC2201903, Grant No. 2021YFC2201901).

\vspace{1cm}

\bibliographystyle{unsrt}
\bibliography{main}

\end{document}